\documentclass[prb,twocolumn,showpacs,preprintnumbers,amsmath,amssymb]{revtex4}

\usepackage{graphicx}
\usepackage[active]{srcltx}
\usepackage{dcolumn}
\usepackage{bm}
\begin{document}
\newcommand{\Bi}{Bi$_2$Sr$_2$CaCu$_2$O$_{8+\delta}$ }

\preprint{APS/123-QED}

\title{Nature of $c$-axis coupling in underdoped
Bi$_2$Sr$_2$CaCu$_2$O$_{8+\delta}$ with varying degrees of disorder}

\author{Panayotis Spathis, Sylvain Colson, Feng Yang, Cornelis J. van der 
Beek}
\affiliation{Laboratoire des Solides Irradi\'{e}s, Ecole 
Polytechnique, CNRS-UMR 7642 \& CEA/DSM/DRECAM, 91128 Palaiseau, France} 
\author{ Piotr Gier{\l}owski}
\affiliation{\mbox{Institute of Physics, Polish Academy 
of Sciences, 32/46 Aleja Lotnik\'ow, 02-668 Warsaw, Poland}}
\author{Takasada Shibauchi, Yuji Matsuda}
\affiliation{\mbox{Department of Physics, Kyoto University, 
Sakyo-ku, Kyoto 606-8502, Japan}}
\author{Marat Gaifullin}
\affiliation{National Institute for Materials Science, 1-2-1 Sengen, Tsukuba, Ibaraki, Japan}
\affiliation{Department of Physics, Loughborough University, Loughborough LE11 3TU, United Kingdom}
\author{Ming Li, Peter H. Kes}
\affiliation{\mbox{Kamerlingh Onnes Laboratorium, 
Rijksuniversiteit Leiden, P.O. Box 9506, 2300 RA Leiden, The 
Netherlands}}

\date{\today}

\begin{abstract}
The dependence of the Josephson Plasma Resonance (JPR) frequency in heavily underdoped 
Bi$_{2}$Sr$_{2}$CaCu$_{2}$O$_{8+\delta}$ on temperature and 
controlled pointlike disorder, introduced by high-energy electron 
irradiation, is cross-correlated and compared to the behavior of the 
$ab$--plane penetration depth. It is found that the zero temperature 
plasma frequency, representative of the superfluid component of the 
$c$-axis spectral weight, decreases proportionally with $T_{c}$ when 
the disorder is increased. The temperature dependence of the JPR 
frequency is the same for all disorder levels, including pristine 
crystals. The reduction of the $c$-axis superfluid density  as 
function of disorder is accounted for by pair-breaking induced by impurity 
scattering in the CuO$_{2}$ planes, rather than by quantum 
fluctuations of the superconducting phase. The reduction of the 
$c$-axis superfluid density as function of temperature follows a 
$T^{2}$--law and is accounted for by quasi-particle hopping through 
impurity induced interlayer states.
\end{abstract}

\maketitle

\section{\label{sec:level1}Introduction}

Significant controversy remains concerning an appropriate model description 
of high temperature superconducting cuprates (HTSC) in the underdoped 
regime, {\em i.e.} the regime in which the number of additional holes per 
Cu, $p$ is smaller than the value $0.16$ at which the critical 
temperature $T_{c}$ is maximum.\cite{Presland91} Whereas it is well established that the charge 
dynamics and transport properties in the normal- and superconducting 
states in the overdoped regime ($ p > 0.16$) are, by and large, determined by 
well-defined quasi-particles, the role of quasi-particles in the 
underdoped regime is debated. The underdoped region of the $(p,T)$ 
phase diagram is characterized by several salient 
features.\cite{Deutscher99} At $T^{*} > T_{c}$, the well-known ``pseudo-gap'' in the 
excitation spectrum opens up. This has been interpreted as either being 
related to the advent of another type of (spin- or charge-) order competing 
with superconductivity, driving $T_{c}$ down as $p$ is diminished, 
or, alternatively, as signalling the formation of precursor Cooper pairs without long-range phase 
coherence. Then, $T_{c}$ is interpreted as the demise of 
long range superconducting phase order due to strong 
thermal\cite{Emery95,Geshkenbein97,Ioffe99} or quantum\cite{Paramekanti2000,Paramekanti2002} phase 
fluctuations. Strong support for this scenario has come from the 
violation of the Glover-Tinkham-Ferrell conductivity sum-rule applied 
to the $c$-axis spectral weight;\cite{Ioffe99} also, the linear relation between $T_{c}$ and the superfluid 
density\cite{Uemura89} has been interpreted as the result of $T_{c}$ 
being determined by phase fluctuations in a Kosterlitz-Thouless-type 
scenario.\cite{Corson99,Kitano2006} A smoking gun for such a scenario 
would be an important reduction of the $c$-axis superfluid density 
$\rho^{c}_{s}$, which is determined by Cooper pair and quasiparticle tunneling 
between adjacent strongly superonducting CuO$_{2}$ layers through the weakly 
superconducting rocksalt-like blocking layers, with respect to the in-plane 
stiffness $\rho^{ab}_{s}$ in 
underdoped cuprates \em below \rm $T_{c}$. 

However, apart from phase fluctuations, other mechanisms for the reduction of $\rho^{c}_{s}$, arising 
from disorder in the crystalline structure of underdoped cuprates, cannot 
be ignored.\cite{Kim2001}  First,  scanning 
tunneling spectroscopy (STS) experiments \cite{Cren2000,Pan2001,Lang2002} 
have revealed large variations of the magnitude of the gap maximum 
$\Delta_{peak}$, as interpreted from conductance curves measured on the 
surface of Bi$_{2}$Sr$_{2}$CaCu$_{2}$O$_{8+\delta}$ crystals. This has  
motivated recent interpretations of weakened $c$-axis superfluid 
response in this material \cite{Singley2004,Shibauchi2006} as well as in  
La$_{2-x}$Sr$_{x}$CuO$_{4}$ \cite{Dordevic2003} in terms of finely 
dispersed 5  -- 20  nm-sized non-superconducting regions within the CuO$_{2}$ 
planes. Such regions may arise from the suppression of the 
superconducting order parameter by dopant atoms, 
\cite{Hettler2000} such as  out-of-plane oxygen atoms in the  
Bi$_{2}$Sr$_{2}$CaCu$_{2}$O$_{8+\delta}$ 
compound.\cite{Nunner2005} 

Moreover, the $d$-wave symmetry of 
the gap function is at the basis of several mechanisms by which 
pointlike disorder reduces the $c$-axis superfluid density. 
The appearance of quasiparticle (virtual-) bound states and their 
smearing by a finite defect density leads to an increase of the density 
of states (DOS) near the nodal $(\pi,\pi)$ directions (the so-called 
lifetime effect).\cite{Hussey2002} The same pointlike defects increase the quasiparticle scattering rate 
$\Gamma_{s}$. It was conjectured that in the case of coherent (in-plane 
momentum preserving)  quasiparticle tunneling, the cancellation of 
these effects leads to disorder-independent low--temperature  $c$-axis 
quasiparticle conductivity and superfluid density.\cite{Latyshev99} 
The approach of Ref.~[\onlinecite{Latyshev99}] neglects the crystal structure of the tetragonal 
HTSC,\cite{Kim2001} which leads to the 
dependence of the interlayer hopping integral $t_{\perp}$ on the in-plane momentum $(k_{x},k_{y})$. For simple tetragonal 
structures, interlayer hopping occurs via Cu $4s$ orbitals in 
adjacent planes. Its momentum dependence $t_{\perp} = t_{\perp}^{0} [ \cos 
k_{x}a - \cos k_{y}a ]^{2}$  is determined by the in--plane overlap of the 
bonding oxygen $2p$ level with the $4s$ level of the neighboring Cu 
atom.\cite{Andersen95,Xiang96} As a result, $c$-axis 
tunneling occurs nearly exclusively for the anti-nodal directions at which quasiparticles 
are unlikely to be excited. In body-centered tetragonal structures such as 
Bi$_{2}$Sr$_{2}$CaCu$_{2}$O$_{8+\delta}$, hopping is also suppressed 
along the $(k_{x},k_{y}) = (\pi,0)$ and $(0,\pi)$ lines, yielding $t_{\perp} = t_{\perp}^{0} [ \cos 
k_{x}a - \cos k_{y}a ]^{2}\cos \frac{1}{2}k_{x}a \cos 
\frac{1}{2}k_{y}a$.\cite{vanderMarel99} In either case, disorder is 
always relevant for the nodal directions. Then, from the lifetime effect, one expects
a quadratic decrease with temperature of the reduced $c$-axis superfluid 
density\cite{Xiang96,Xiang98}
\begin{equation}
    \frac{\rho^{c}_{s}(T)}{\rho^{c}_{s}(0)} \propto  1  - \alpha_{c}
    \frac{8 \pi}{3}\frac{\Gamma_{s}}{\Delta_{0}}\left( \frac{T}{\Delta_{0}} 
    \right)^{2}, \hspace{0.7cm} ( k_{B}T \ll \Gamma_{s}  ) .
 \label{eq:lifetime}
 \end{equation}
Here $\alpha_{c}$ is a dimensionless constant of order unity and 
the parameter $\Delta_{0}$ was assumed, in Refs.~\onlinecite{Xiang96} 
and \onlinecite{Xiang98}, to correspond to the maximum 
amplitude of the Bardeen-Cooper-Schrieffer $d$-wave gap. Eq.~(\ref{eq:lifetime}) 
essentially differs from that derived for the $ab$-plane superfluid 
density
\begin{equation} 
     \frac{\rho^{ab}_{s}(T)}{\rho^{ab}_{s}(0)} \propto 1 - 
     \alpha_{ab} \frac{\Delta_{0}}{\Gamma_{s}} \left( \frac{T}{\Delta_{0}} 
    \right)^{2}, \hspace{0.7cm} ( k_{B}T \ll \Gamma_{s}  ) 
     \label{eq:lifetime-ab}
\end{equation}
in that the leading temperature--dependent term has a coefficient that 
is smaller by a factor $(\Gamma_{s}/\Delta_{0})^{2}$.\cite{Xiang96} 
The presence of defects in the rocksalt-like (BiO) layers tends to break the $d$-wave symmetry  of the 
hopping integral, and renders quasiparticle hopping possible for other values of the in-plane 
momentum, and notably along the order parameter 
nodes.\cite{Xiang96,Xiang98,Kim98,Kim2001} A condition for this 
``impurity-assisted hopping'' (IAH) to be effective is an 
anisotropic scattering matrix of the interplane 
defects. Notably, for strong forward scattering, the 
result 
\begin{equation}
    \rho^{c}_{s}(T) \approx 2 \pi V_{1} \Delta_{0} N^{2}(E_{F})\left[1  - 8 \ln 2 \left( 
    \frac{T}{\Delta_{0}} \right)^{2} \right],
\label{eq:IAH}
\end{equation}
was obtained for $ \Gamma_{s} \ll k_{B}T \ll \frac{1}{2}\left[ 
2 \pi V_{1} \Delta_{0} N(E_{F})/(t_{\perp}^{0 })^{2}\right]^{1/3}T_{c}$.
\cite{Radtke96,Xiang96} Here $V_{1}$ is the magnitude of the impurity 
scattering potential of the out-of-plane defects and $N(E_{F})$ is the density of states at the  
Fermi level in the normal state. The effect of impurities can be distinguished from that of 
boson-assisted interlayer hopping; for the latter, a very similar result is obtained, but with the leading 
temperature--dependent term proportional to $T^{3}$.\cite{Hirschfeld97,Xiang98} 
Finally, direct hopping of quasiparticles was suggested to lead to  
a small, linearly temperature-dependent, reduction of $\rho^{c}_{s}$.\cite{Radtke96}


In this paper, we address the mechanism by which the $c$-axis 
superfluid density in underdoped 
Bi$_{2}$Sr$_{2}$CaCu$_{2}$O$_{8+\delta}$ ( with $p = 0.10$ ) is 
reduced by using disorder, in the form of Frenkel pairs introduced by 
high energy electron irradiation, as an independent control 
parameter. Electron irradiation, the effects of which are  taken to be
similar to those of Zn-doping,\cite{Albenque2000} has previously been 
used to study the effect of pointlike disorder on 
the resistivity, critical temperature,\cite{Albenque2003} and Nernst effect of YBa$_{2}$Cu$_{3}$O$_{7}$ and 
YBa$_{2}$Cu$_{3}$O$_{6.6}$.\cite{Albenque2006} In the latter 
material, electron irradiation eventually leads (at high fluences) to the breakdown of 
the well-known Abrikosov Gor'kov relation\cite{Abrikosov60,Sun95} 
\begin{equation}
\ln\left( \frac{T_{c}}{T_{c0}} \right) = 
\Psi\left(\frac{1}{2}\right) - \Psi \left( \frac{1}{2} + 
\frac{\Gamma}{2  \pi k_{B} T_{c} } \right)
\label{eq:AG}    
\end{equation} 
(with $T_{c0}$ the critical temperature when the normal state scattering 
rate $\Gamma$ is equal to zero,  and $\Psi$ the digamma function) as well as a significant 
increase of the fluctuation regime near $T_{c}$.\cite{Albenque2006}  Both effects were interpreted as 
the effect of strong superconducting phase 
fluctuations.\cite{Albenque2003,Albenque2006}  The in-plane and $c$-axis superfluid 
densities of Zn-doped YBa$_{2}$Cu$_{3}$O$_{7-\delta}$ were studied by 
Panagopoulos {\em et al.} \cite{Panagopoulos96} and by Fukuzumi, 
Mizuhashi, and Uchida.\cite{Fukuzumi2000} The progressive inclusion of Zn leads to a rapid decrease of the in-plane 
superfluid density $\rho_{s}^{ab}$, corresponding to an increase of 
the in-plane penetration depth $\lambda_{ab}(0) \propto
(\rho_{s}^{ab})^{-1/2}$, and a more modest decrease of 
$\rho_{s}^{c} \propto \sigma_{c}(T_{c}) \propto T_{c}$, that violates the 
$c$-axis conductivity sum-rule \cite{Ioffe99,Fukuzumi2000} [$\sigma_{c}(T_{c})$ is the 
$c$-axis conductivity at $T_{c}$]. As for the low--$T$ 
temperature dependence, a gradual change of both $\rho_{s}^{ab}$ and $\rho_{s}^{c}$ 
from $T$--linear to $T$--squared has been reported.\cite{Panagopoulos96}
Studies on Bi$_{2}$Sr$_{2}$CaCu$_{2}$O$_{8+\delta}$ are limited to 
electron irradiation of the single crystalline optimally doped 
material, that show a linear decrease of $T_{c}$ with electron 
fluence.\cite{Albenque95,Behnia2000,Nakamae2001} 
The $c$-axis superfluid density in a  underdoped pristine
Bi$_{2}$Sr$_{2}$CaCu$_{2}$O$_{8+\delta}$ single crystal has been previously 
studied by Gaifullin {\em et al.}, who invoked the IAH model to 
explain the much stronger temperature dependence of $\rho_{s}^{c}$ in 
underdoped with respect to optimally doped 
Bi$_{2}$Sr$_{2}$CaCu$_{2}$O$_{8+\delta}$.\cite{Gaifullin99}

Below, we report on $c$-axis coupling in the superconducting state 
measured through the Josephson Plasma Resonance 
(JPR),\cite{Matsuda95,Matsuda97,Shibauchi99,Gaifullin2000} which, in our underdoped 
Bi$_{2}$Sr$_{2}$CaCu$_{2}$O$_{8+\delta}$ crystals takes place in the 
microwave frequency regime below 70 GHz.  The JPR
frequency $f_{pl}$ is sensitive to the value of $\Delta_{0}$, as well as 
to fluctuations of the superconducting order parameter phase in the 
CuO$_{2}$ planes.\cite{Paramekanti2000} The evolution of $f_{pl}(T)$ with temperature depends simultaneously on the quasiparticle 
dynamics and on the strength of fluctuations;  the plasma resonance 
peak is broadened both by the quasiparticle tunneling rate and by 
crystalline disorder.\cite{Dordevic2003,Koshelev99} However, the 
dependence of $f_{pl}^{2}\propto  \rho_{s}^{c}$ on the disorder strength is expected to be 
quite different, depending on which mechanism is predominant. In the 
following, we show that the disorder dependence of the $c$-axis 
plasma frequency is a sensitive probe, that allows one to identify in 
detail what physical mechanisms are at the basis of the reduction of 
the superfluid density in Bi$_{2}$Sr$_{2}$CaCu$_{2}$O$_{8+\delta}$. It 
turns out that, even in our heavily underdoped crystals, 
(incoherent) $c$-axis quasiparticle hopping is essential for a consistent description of 
the data. We find that the energy scale $\Delta_{0}$, which turns out to be 
$Delta_{0} \approx 2.5 k_{B}T_{c}$ for all underdoped crystals, is to be 
interpreted as an energy scale governing nodal quasi-particle 
excitations.

\begin{figure}[b]
\begin{center}
\includegraphics[width=8cm,keepaspectratio]{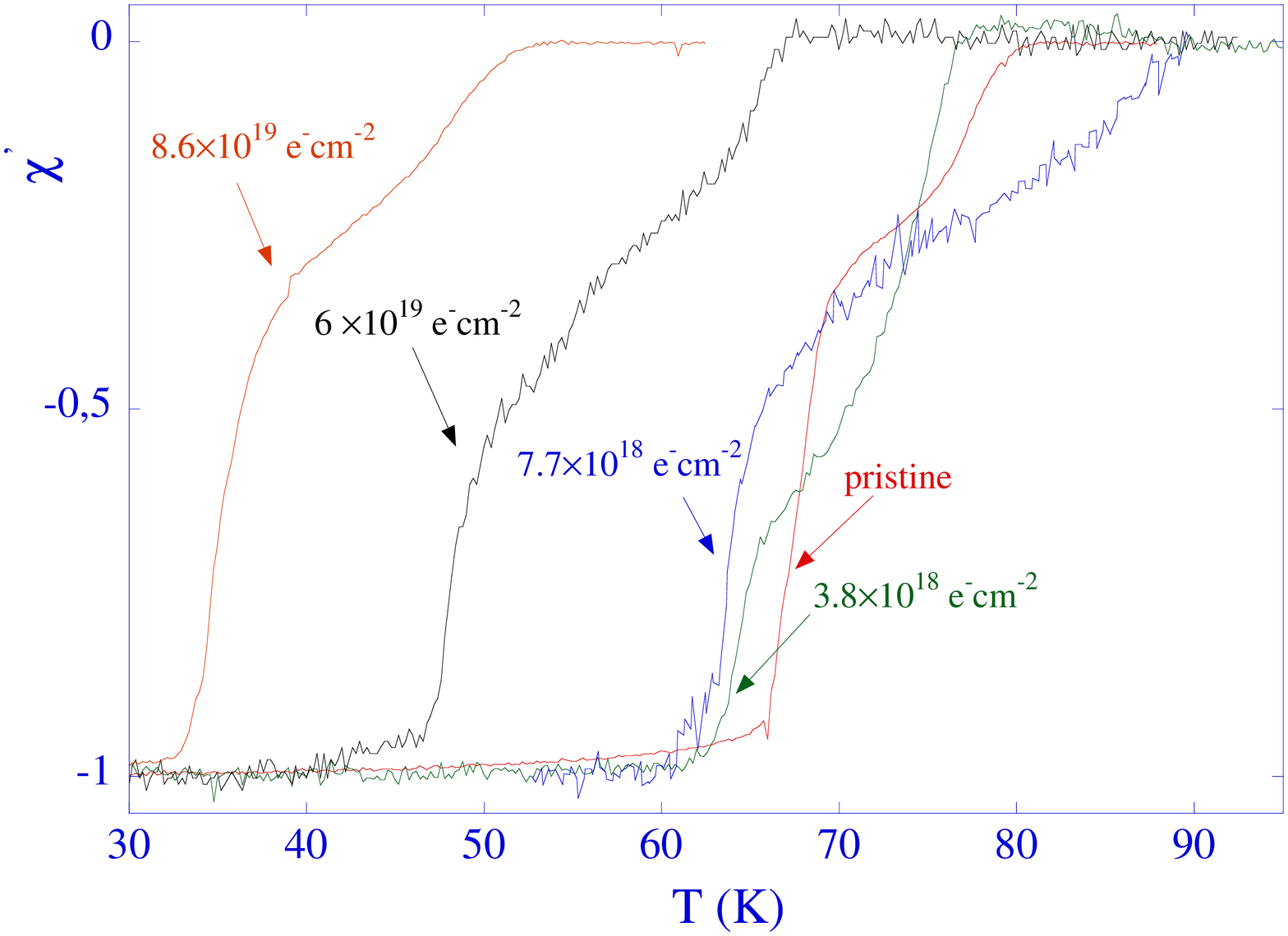}
\caption{ (color online) Real part of the ac magnetic susceptibility 
of the investigated underdoped Bi$_{2}$Sr$_{2}$CaCu$_{2}$O$_{8+\delta}$ 
crystals. The crystals were irradiated at 22 K with 2.3 MeV electrons 
to the indicated fluences. The ac field amplitude $h_{ac}=4.2$ mOe, 
the ac frequency was 560 Hz. The curves show a screening onset determined by 
a surface layer containing optimally doped material; this screening 
is suppressed when $h_{ac} \gtrsim 0.5$ Oe. The steep drop at the 
lower end of the transitions corresponds to bulk screening by the 
underdoped material.}
\label{fig:ac-susceptibility}
\end{center}
\end{figure}

\section{\label{sec:level2}Experimental Details}

The underdoped ($T_{c}=65 \pm 0.5$ K, $p \approx 0.10$) Bi$_{2}$Sr$_{2}$CaCu$_{2}$O$_{8+\delta}$ 
single crystals, of typical dimensions $500 \times 300 \times 40 
\mu$m$^{3}$, were selected from the same boule, grown by 
the travelling solvent floating zone method at the FOM-ALMOS center, 
the Netherlands, in 25 mBar O$_2$ partial pressure.\cite{MingLi2002} The 
crystals were annealed for one week in flowing N$_2$ gas. We have 
also measured a set of optimally doped control samples ( $T_{c} = 86$ 
K ). These were also grown by the  travelling solvent floating zone technique, at 
200 mbar oxygen partial pressure, and subsequently annealed in air at 
800$^{\circ}$C. The crystals were irradiated with 2.3 MeV electrons using 
the Van de Graaff accelerator 
at the Laboratoire des Solides Irradi\'{e}s. The beam was directed 
along the crystalline $c$-axis during the irradiation. To prevent recombination 
and clustering of point defects, the irradiation is carried out with 
the crystals immersed in a liquid hydrogen bath ( 22 K ). The 
electron flux is limited to $2\times 10^{14}$ e$^{-}$cm$^{-2}$ per second. 
Crystals UD5-UD8 were irradiated to a total fluence of $0.53 \times 
10^{18}$, $3 \times 10^{18}$, $7.7 \times 10^{18}$, 
and $8.8 \times 10^{19}$ e$^{-}$cm$^{-2}$ respectively. After 
measurements, crystal UD5 was irradiated a second time to a total fluence of $6.0 \times 
10^{19}$ e$^{-}$cm$^{-2}$ and was henceforth labeled UD5b. The high energy 
electron irradiation creates random 
atomic displacements in the form of Frenkel pairs, both in the 
CuO$_2$ bilayers and in the intermediate cation layers, throughout the 
samples.

\begin{figure}[t]
\begin{center}
\includegraphics[width=8cm,keepaspectratio]{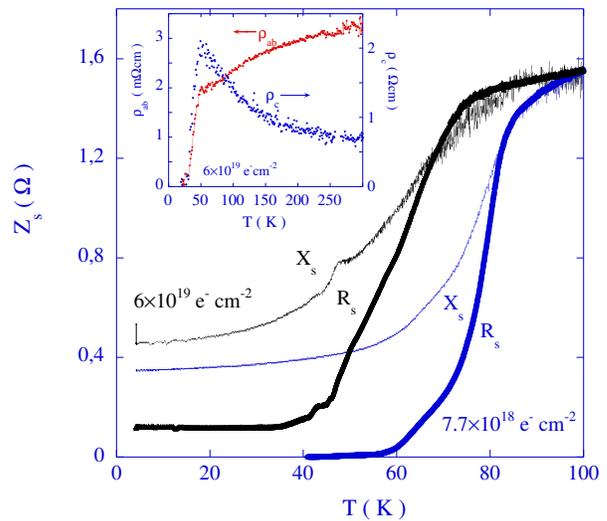}
\caption{ (color online) Real ( $R_{s}$ ) and imaginary ( $X_{s}$ ) 
parts of the surface impedance $Z_{s}$ of underdoped 
Bi$_{2}$Sr$_{2}$CaCu$_{2}$O$_{8+\delta}$ crystals, irradiated 
with $7.7 \times 10^{18}$ and $6\times10^{19}$ electrons cm$^{-2}$, 
respectively.  The data were obtained from the resonance frequency 
shift and the quality factor of a superconducting Pb cavity operated 
in the TE$_{011}$ mode. The inset shows the $ab$--plane-- and the $c$-axis dc resistivity of the crystal irradiated with 
$6\times10^{19}$ electrons cm$^{-2}$.}
\label{fig:resistive-plot}
\end{center}
\end{figure}

\begin{figure}[t]
\begin{center}
\includegraphics[width=8cm,keepaspectratio]{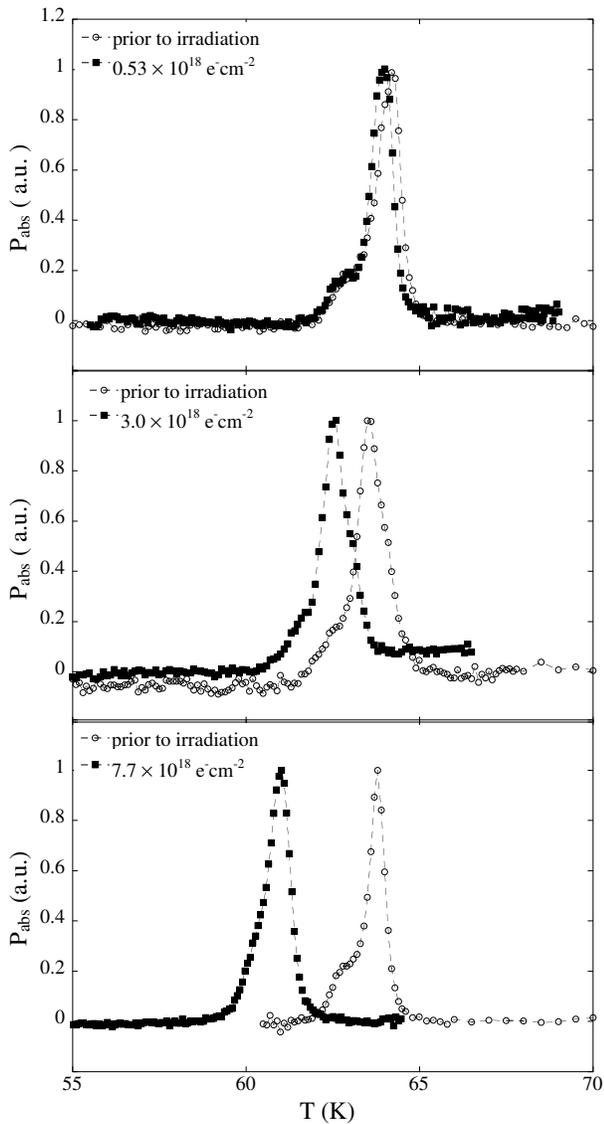}
\caption{Microwave absorption, measured using the 
TM$_{010}$ mode ( 19.2 GHz ) of one  of the OFHC copper cavities, of three of the underdoped 
Bi$_{2}$Sr$_{2}$CaCu$_{2}$O$_{8+\delta}$ crystals before and after 
irradiation with $5.3\times 10^{17}$, $3\times 10^{18}$, and 
$7.7\times10^{18}$ 2.3 MeV electrons cm$^{-2}$, respectively.}
\label{fig:MW-vs-T}
\end{center}
\end{figure}

The superconducting transition temperature $T_{c}$ was determined 
by ac susceptibility measurements using a driving field of amplitude 4.2 mOe 
and a frequency of 560 Hz, directed parallel to the $c$-axis. For all 
underdoped crystals, the superconducting transition is rather broad. 
The superconducting transition takes place in two steps:  there is a slow increase of screening at high temperature, 
followed by a rapid step of the diamagnetic signal at lower 
temperature ( Figure \ref{fig:ac-susceptibility}  ). The high temperature screening 
vanishes when the excitation field amplitude is increased beyond 
 0.5 Oe, while the step at lower temperature is robust. This shows that doping is macroscopically inhomogeneous, and that the 
crystals are surrounded by a thin surface layer of higher doping. 
This layer could not be eliminated by cutting the crystals. 
The overall shape of the transition is unaffected by the electron irradiation. The 
transition widths are of the order of 4 K, which is usual for such 
low doping. After irradiation, the transition widths slightly 
increase. For all crystals, the transition to zero dc resistivity and 
bulk superconductivity occurs at the temperature at which the lower 
screening step takes place, see {\em e.g.} the Inset to 
Fig.~\ref{fig:resistive-plot}. Therefore, the lower temperature feature was adopted as 
characterizing the bulk $T_{c}$ of the underdoped crystals. 


Crystals were further characterized by the measurement of the temperature variation of the in-plane penetration depth, 
$\lambda_{ab}(T)/\lambda_{ab}(0) - 1 \equiv \Delta \lambda_{ab} / 
\lambda_{ab}$. For this, a crystal is mounted on a sapphire rod, 
in the center of a superconducting (Pb) resonant cavity immersed in 
liquid $^{4}$He, and operated in the TE$_{011}$ mode. The cavity 
resonant frequency was $f \sim 27.8 $ GHz, and the quality factor  
$Q \sim 4\times 10^5$. 
The crystal is mounted in such a way that the magnetic microwave field is perpendicular to its $ab$ plane
and solely in-plane screening currents are induced. From the shift $\Delta 
f$ of the cavity resonance frequency induced by the sample, we 
determine the surface reactance $X_{s} = 2 \pi \mu_{0} G_{2} \Delta 
f$ (with $\mu_{0} = 4\pi \times 10^{-7}$ Hm$^{-1}$). The surface 
resistance $R_{s} = 2 \pi \mu_{0} G_{1} (\Delta Q)^{-1}$ was obtained 
from the change of the quality factor. The geometrical factors $G_{1}$ and $G_{2}$ were determined 
by comparing the surface impedance in the normal state, $X_{s} = 
R_{s} = \pi \mu_{0} f \delta_{s}$, to the value 
expected from the normal state resistivity, $\rho =  \pi 
\mu_{0} f \delta_{s}^{2}$.\cite{Shibauchi94} 
It was retrospectively checked that all measurements were carried out in the skin effect 
regime, in which the normal state skin depth $\delta_{s}$ is much smaller 
than the sample dimensions. The relative change of the penetration 
depth was determined from the behavior of the surface reactance at 
low temperature, at which $X_{s} \approx 2 \pi \mu_{0} f \lambda_{ab}$. 

Figure~\ref{fig:resistive-plot} shows that the temperature 
at which the decrease of the surface resistance $R_{s}$ was observed 
corresponds to the (high temperature) onset of screening in the ac susceptibility measurement. 
This indicates that the surface skin depth of thickness $\sim 7 \mu$m probed by 
the microwave field contains patches with larger hole content $p$. 

%
%
%

\begin{figure}[b]
\begin{center}
\includegraphics[width=8cm,keepaspectratio]{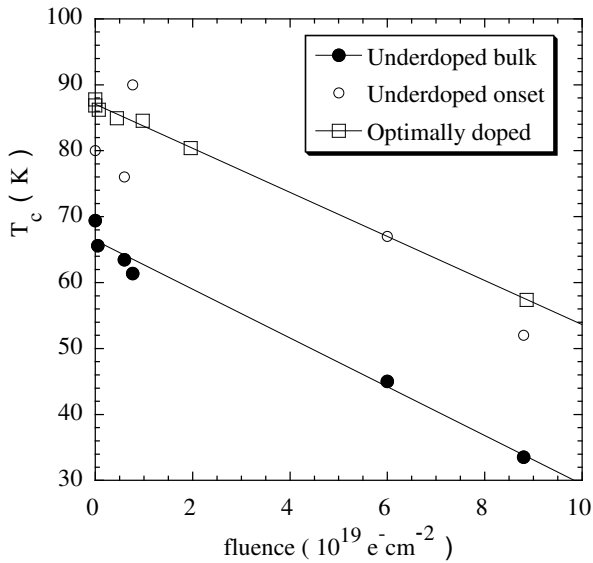}
\caption{ Variation of the critical temperature with 
electron fluence for single-crystalline  underdoped 
Bi$_{2}$Sr$_{2}$CaCu$_{2}$O$_{8+\delta}$ ($\bullet$).  A comparison with the $T_{c}$ 
of a series of optimally doped control crystals ($\Box$) shows that 
the onset temperature of magnetic screening ($\circ$) corresponds 
to flux exclusion by a surface region containing optimally doped material. }
\label{fig:Tc-vs-dose}
\end{center}
\end{figure}

The JPR measurements were performed using the cavity perturbation technique, 
using the TM$_{01n}$ modes ($n=0,...,4$) of two Oxygen-Free High 
Conductivity (OFHC) copper resonant cavities ( $Q\sim 4000-10000$ ), mounted 
on a cryocooler cold head. The measurement 
frequencies ranged between 19.2 and 39.6 GHz.\cite{Colson2003} Further 
measurements were made applying the bolometric technique, using 
waveguides in the TE$_{01}$ travelling wave mode.\cite{Gaifullin99} 
In both measurement set-ups, the electrical microwave field is applied along the $c$-axis 
of the crystal. In contrast to the previously described surface impedance measurements, screening of the 
electric microwave field is very poor because of the high electronic anisotropy of the 
crystals. The underdoped samples are in the complete depolarization 
regime and thus the bulk electromagnetic response is probed. By 
monitoring the power absorption as a function or frequency for a 
fixed temperature, the Josephson Plasma Resonance is detected as a sharp absorption peak in 
the microwave response, see Fig.~\ref{fig:MW-vs-T}. We determine the 
JPR frequency at a given temperature, $f_{pl}(T)$, as the measurement 
frequency at the temperature at which dissipation is maximum.

\begin{figure}[t]
\begin{center}
\includegraphics[width=7cm,keepaspectratio]{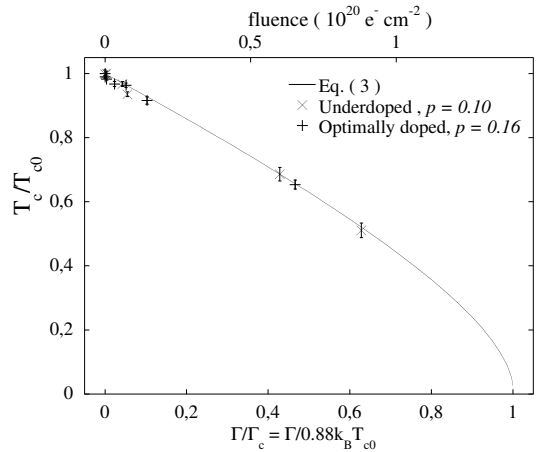}
\caption{ Critical temperatures $T_{c}$ 
normalized to the critical temperature $T_{c0}$ of the unirradiated 
crystals, and superimposed on the Abrikosov-Gor'kov relation 
(\protect\ref{eq:AG}). Note that the actual data points should be 
shifted downwards along the curve, because $T_{c0}$ does not truly
correspond to the critical temperature in the absence of disorder.}
\label{fig:AG}
\end{center}
\end{figure}

\section{\label{sec:level3}Results}

Figure~\ref{fig:Tc-vs-dose} collects the values of the critical 
temperature as function of electron dose, for the set of underdoped 
samples as well as the optimally doped control samples. Both the 
$T_{c}$ of the underdoped and the optimally doped crystals decrease
linearly with irradiation fluence. The derivative of $T_{c}$ with 
respect to fluence of the optimally doped crystals concurs with that measured 
by Behnia {\em et al.}\cite{Behnia2000} and Nakamae {\em et al.}\cite{Nakamae2001} but is two 
times lower than that measured by Rullier-Albenque {\em et  
al.}.\cite{Albenque95} The overlap between the variation with fluence 
of the screening onset temperature in the underdoped crystals and the 
$T_{c}$ of the optimally doped crystals shows that the thin surface 
layer on the underdoped Bi$_{2}$Sr$_{2}$CaCu$_{2}$O$_{8+\delta}$  has $p \sim 0.16$.
The critical temperatures, normalized to the critical temperatures 
$T_{c0}$ of the unirradiated crystals, can be superimposed on 
Eq.~(\ref{eq:AG}), yielding estimates of the normal state 
scattering rate $\Gamma$ (see Fig.~\ref{fig:AG}). 
This procedure supposes that $T_{c0}$ corresponds to the critical 
temperature in the absence of disorder; we shall see below that this 
is not justified, so that the estimated $\Gamma$ values are in fact 
lower limits for each crystal.

\begin{figure}[t]
\begin{center}
\includegraphics[width=8cm,keepaspectratio]{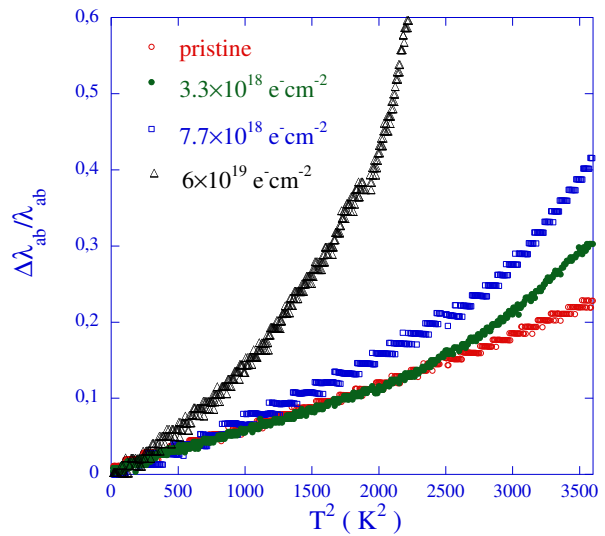}
\caption{ (color online) Relative change of the $ab$--plane 
penetration depth $\Delta \lambda_{ab} / \lambda_{ab} (0) =  [ X_{s}(T)/X_{s}(T \rightarrow 
0) - 1]/ 2 \pi \mu_{0} f $, as obtained 
from the change in the inductive part $X_{s}$ of the surface 
impedance. }
\label{fig:delta-lambda-plot}
\end{center}
\end{figure}

The relative change with temperature of the in-plane penetration 
depth $\lambda_{ab}$ is depicted in Fig.~\ref{fig:delta-lambda-plot}. 
For all underdoped crystals, including the unirradiated ones, $\lambda_{ab}(T) - \lambda_{ab}(0)$ varies 
quadratically with temperature at low $T$. Such a temperature dependence has been 
associated with quasiparticle scattering in the unitary limit by point defects 
situated within the CuO$_{2}$ planes of the $d$-wave 
superconductor, {\em i.e.} $N(E_{F})V \gg 1$, $\Gamma \sim n_{d}/ \pi 
N(E_{F})$, and $\Gamma_s \sim 0.6 (\Gamma \Delta_0 )^{1/2}$. \cite{Preosti94} 
Here  $V$ is the scattering potential of the defects \em in \rm the planes, 
with density  $n_{d}$. The quadratic temperature dependence 
of $\Delta \lambda_{ab}(T)$ is at odds with a possible important role of thermal phase 
fluctuations, for which a linear $T$--dependence was predicted.\cite{Roddick95}
As for the magnitude of the $T^{2}$-contribution to $\lambda_{ab}$, 
a very modest change is found for the lower irradiation fluences. 
Only for fluences exceeding $10^{19}$ e$^{-}$cm$^{-2}$ does the 
in-plane penetration depth increase significantly with defect density.

\begin{figure}[t]
\begin{center}
\includegraphics[width=8cm,keepaspectratio]{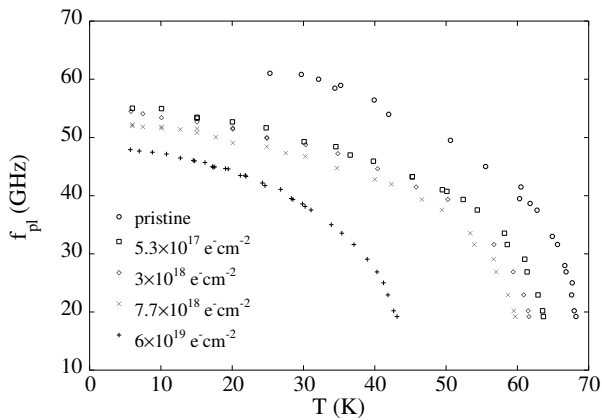}
\caption{Temperature dependence of the JPR frequency of 
electron-irradiated underdoped Bi$_{2}$Sr$_{2}$CaCu$_{2}$O$_{8+\delta}$.}
\label{fig:Fp-vs-T}
\end{center}
\end{figure}

We now switch to the central results concerning the Josephson Plasma 
Resonance. Figure \ref{fig:Fp-vs-T} shows the JPR frequency 
$f_{pl}(T)$ of crystals UD5-UD7 as function of temperature ( measured 
in Earth's magnetic field ). The temperature at which $f_{pl}(T)$ 
extrapolates to zero is well-defined and corresponds to the 
critical temperature of the bulk, underdoped portion of the crystals, 
\em i.e. \rm  the main transition in the ac susceptibility and  zero resistance. This shows that 
the JPR probes the $c$-axis response in the heavily 
underdoped bulk and is insensitive to the surface quality of the samples. 
From Fig.~\ref{fig:Fp-vs-T} one sees that not only $T_c$, but also $f_{pl}$ is strongly depressed 
by the electron irradiation. Fig.~\ref{fig:jc-vs-Tc} 
collects values of the low-temperature extrapolated value $f_{pl}(0)$ versus the critical temperature, and reveals the 
proportionality between  $f_{pl}^{2}(0)$ and $T_{c}$. This dependence 
is clearly different from the variation of $f_{pl}$ with oxygen doping. 
The same Figure recapitulates results for doping levels $p = 0.13$  
\cite{Gaifullin99} and 0.11,\cite{Colson2003} the evolution of which 
recalls the exponential $f_{pl}^{2}(0)(T_{c})$--dependence found by 
Shibauchi {\em et al.}.\cite{Shibauchi2006} The results are somewhat 
similar to those obtained by Fukuzumi {\em et al.} for Zn-doped and 
oxygen-deficient YBa$_{2}$Cu$_{3}$O$_{7-\delta}$:\cite{Fukuzumi2000} 
however, their Fig.~5 shows a linear dependence of $f_{pl}$ on $T_{c}$ for 
underdoped YBa$_{2}$(Cu$_{1-x}$Zn$_{x}$)$_{3}$O$_{6.63}$, and a 
hyperbolic or exponential evolution of $f_{pl}(T_{c})$ as function of $\delta$.

As for the temperature variation of $f_{pl}^{2}(T)\propto \rho_{s}^c$, 
it turns out to be identical for all the underdoped crystals, including the pristine ones, and does not depend on the 
defect density. Fig.~(\ref{fig:fp-scaled}) shows $f_{pl}^2(T)/f_{pl}^2(0)$ plotted versus 
the reduced temperature $T/T_{c}$. For $h f_{pl} < \Gamma$, 
this is representative of the $c$-axis superfluid fraction 
$\rho_s^c$. The same graph may also be interpreted as the maximum 
Josephson ($c$-axis) critical current, 
\begin{equation}
    j_{c}^{c} = 2 \pi (h/2e) \epsilon \epsilon_{0} s^{-1} f_{pl}^{2} = 
    \frac{h}{ 4 \pi e \mu_{0} \lambda_{c}^{2} s}
\end{equation}
normalized to its value for $T\rightarrow 0$. Here $s$ is the spacing between 
CuO$_{2}$ planes, the $c$-axis dielectric constant $\epsilon 
\approx 11.5$,\cite{Gaifullin2001} and $\epsilon_{0} = 
8.854\times10^{-12}$ Fm$^{-1}$.

\section{Discussion}

To distinguish between the different mechanisms responsible for the 
reduction of the $c$-axis plasma frequency as temperature is 
increased, we dispose of three tools. First, there is the variation of 
$f_{pl}^{2}(0)$ with disorder strength, which manifests itself 
starting from the smallest electron fluences, and proportionally 
follows the evolution of $T_{c}$ with disorder. Second is the temperature variation 
$f_{pl}^{2}(T)/f_{pl}^{2}(0)$, that follows a $1 - a 
(T/T_{c})^{\alpha}$ dependence, with $\alpha \sim 2$ 
independent of the disorder strength. Finally, there is the comparison with the behavior of 
the in-plane penetration depth, $\lambda_{ab}(T)/\lambda_{ab}(0) \sim 
1 + \beta T^{2}$. A successful model description should account for all three dependences correctly.

\begin{figure}[t]
\begin{center}
\includegraphics[width=8cm,keepaspectratio]{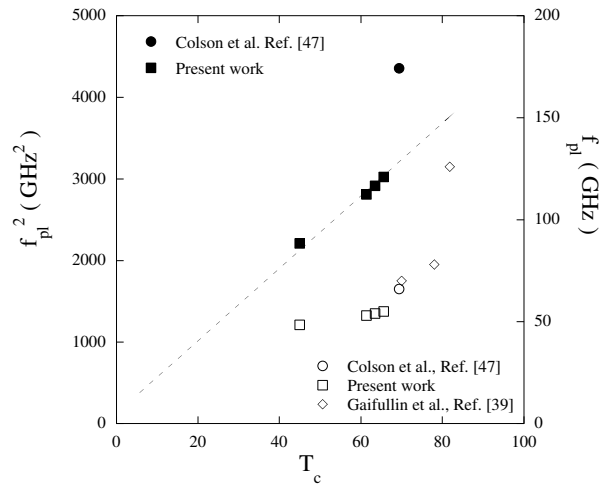}
\caption{$f_{pl}(0)$ versus $T_{c}$ (open symbols), 
as well as $f_{pl}^{2}(0) \propto \rho_{s}^{c}(0) \propto j_{c}^{c}(0)$ 
versus $T_{c}$ (closed symbols), for both the underdoped 
irradiated crystals (squares) and  a set of crystals with different doping 
levels ($p \approx 0.13$,\protect\cite{Gaifullin99} and $p \approx 
0.11$\protect\cite{Colson2003}).}
\label{fig:jc-vs-Tc}
\end{center}
\end{figure}

The theoretically expected low-temperature dependence $\lambda_{c}(T)$ depends 
strongly on a number of circumstances. First is the question whether
superconductive coupling is three-dimensional ( \em i.e. \rm the $c$-axis 
momentum $k_z$ is a good quantum number ),\cite{Xiang96} or whether it is mediated
by Josephson tunneling between CuO$_{2}$ layers. Josephson coupling 
can be weakened by direct\cite{Radtke96,Latyshev99} or
boson-assisted quasiparticle tunneling,\cite{Radtke96,Xiang98,Hirschfeld97} 
or by tunneling that involves intermediate defect-induced states between the 
layers (IAH).\cite{Radtke96,Xiang96} For ``direct'' tunneling, \em 
e.g. \rm occurring through direct overlap of the superconducting wave functions 
in the CuO$_{2}$ planes, one distinguishes the case of conserved in-plane momentum 
$k_\parallel$ (``coherent'' tunneling) from the case where it is not 
preserved (``incoherent'' tunneling \cite{Latyshev99} --- this 
situation would yield a vanishing $j_{c}^{c}$ in a $d$--wave superconductor). A 
prevailing effect of nodal quasiparticles leads to a more rapid 
decrease of $j_{c}^{c}$ , see Eq.~(\ref{eq:lifetime}).
Finally, in all cases the anisotropy of the transfer matrix $t_{\perp}$ 
is expected to have an important influence on the temperature 
dependence of $\lambda_{c}$,\cite{Xiang96,Maki2003} notably 
reconciling a weak $T$--dependence of $\lambda_{c}$ with a strong
variation of $\lambda_{ab}(T)$. Here, the experimentally observed temperature dependence of 
$f_{pl}^{2}(T)/f_{pl}^{2}(0)$ actually allows one to discard a 
dominant role of a possible $d$-wave symmetry of the transfer matrix 
$t_{\perp}$,\cite{Xiang96} since ( for $ k_{B}T \gg \hbar\Gamma_{s}$ ) this leads to 
a weak temperature dependence, $f_{pl}^{2}(T)/f_{pl}^{2}(0) \sim 
1 - \tilde{a} T^{5}$, observed in slightly overdoped  
Bi$_{2}$Sr$_{2}$CaCu$_{2}$O$_{8+\delta}$\cite{Gaifullin99,Gaifullin2001} 
and optimally doped HgBa$_2$Ce$_{n-1}$Cu$_n$O$_{2n+2+\delta}$,\cite{Panagopoulos97} but 
not in the present data on underdoped Bi$_{2}$Sr$_{2}$CaCu$_{2}$O$_{8+\delta}$. 
The modification of $t_{\perp}$ arising from the body-centered Bravais 
lattice of the Bi$_{2}$Sr$_{2}$CaCu$_{2}$O$_{8+\delta}$ compound will 
influence the maximum $c$-axis critical current. However, it will not change 
the expected $1 - \tilde{a} T^{5}$ temperature dependence, since this arises 
from the specific coincidence in $k$-space of the zero of $t_{\perp}(k_{x},k_{y})$ with 
the nodal direction of the order parameter. Thus, models for superconductive 
coupling,\cite{Xiang96} or direct Josephson coupling with a vanishing 
hopping integral along the nodal line\cite{Maki2003} are in 
inadequacy with the data. 

We next exclude a dominant role of direct quasiparticle tunneling. Even 
though the similar $T^{2}$--dependences of the low--temperature 
$ab$--plane and $c$-axis penetration depths suggests such coupling, 
the disorder dependence is at odds with the experimental result. 
Radtke {\em et al.}\cite{Radtke96} and Latyshev {\em et 
al.}\cite{Latyshev99} find that for direct coupling, the low temperature 
$c$-axis critical current  
\begin{equation}
    j_{c}^{c,direct} = \frac{\pi\sigma_{q}^{c}(0)\Delta_{0}}{e s} = 
    \frac{4 \pi e t_{\perp}^{2}N_{n}(E_{F})}{h} 
\end{equation}
is, for $k_{B}T \ll \hbar \Gamma_{s} \sim 20-30$ K, to lowest order independent of the defect 
density due to the cancellation of the scattering-rate dependences 
of the quasi-particle conductivity $\sigma_{q}^{c}$ and $\Delta_{0}$. 
The model was further worked out by Kim and Carbotte, who find 
that to first order 
\begin{equation}
    j_{c}^{c,direct} \propto 1 - \alpha 
    \frac{\Gamma_{s}}{\sqrt{\Gamma_{s}^{2} + \Delta_{0}^{2}}} \sim  1 - 
    \alpha \frac{1}{\sqrt{1+ \Delta_{0}/\Gamma}}
    \label{eq:Wonkee}
    \end{equation}
both for the case of constant $t_{\perp}$ (where $\alpha \approx 1$) and angular--dependent 
$t_{\perp}$ ($\alpha = \frac{16}{9}$).\cite{Kim2001} The nonlinear dependence (\ref{eq:Wonkee}) is at odds 
with the observed linear evolution of $j_{c}^{c}$ with irradiation fluence. 

The temperature and disorder-dependence of the low temperature 
$c$-axis JPR frequency is more successfully described by a 
model for incoherent tunneling. According to Latyshev {\em et al.}, an 
incoherent tunneling process yields $j_{c}^{c,i} \approx j_{c}^{c,direct} 
\Delta_{0}/E_{F} \approx 4 \pi e t_{\perp}^{2}N_{n}(E_{F})\Delta_{0}/ 
h E_{F}$. The extra factor $\Delta_{0}$ then explains the 
linear relation between $j_{c}^{c}(0)$ and $T_{c}$, accepting that in 
a $d$-wave superconductor with impurity scattering in the unitary 
limit, $\Delta_{0}(\Gamma)$ is simply proportional to 
$T_{c}(\Gamma)$.\cite{Sun95} The linear dependence on $\Delta_{0}$ is found in the IAH 
model, see Eq.~(\ref{eq:IAH}).\cite{Radtke96,Xiang96,Hirschfeld97} The 
latter  expression consistently describes the fact that the temperature 
dependence of $f_{pl}$ does not change with defect density: the 
temperature dependent term writes $(T/\Delta_{0})^{2} \propto 
(T/T_{c})^{2}$. For the same reason, the ``lifetime effect'', 
Eq.~(\ref{eq:lifetime}), does not describe the data: in the unitary 
limit, the leading temperature--dependent term has an extra factor 
$\Gamma_{s}/\Delta_{0} \sim ( \Gamma / \Delta_{0} )^{1/2} $ 
and is therefore expected to strongly depend on defect density. The observed 
defect--density independence of $\rho_{s}^{c}(T)/\rho_{s}^{c}$ would 
require the strength of the scattering potential of the individual 
irradiation defects in the CuO$_{2}$ planes to be in the Born limit, which contradicts the 
results on the temperature dependence of the $ab$--plane penetration 
depth. We note that the toy model for incoherent hopping of 
Ref.~[\onlinecite{Kim2001}], which yields $\lambda_{c}^{-2} \propto [1 - \frac{5}{14} 
(\Gamma_{s}/\Delta_{0})^{2} - \ldots ] \sim [ 1 - \frac{3}{14} 
(\Gamma/\Delta_{0}) - \ldots ]$, also describes the initial linear 
decrease of $j_{c}^{c}(T_{c})$ (Eq.~(31) of 
Ref.~[\onlinecite{Kim2001}]).

\begin{figure}[t]
\begin{center}
\includegraphics[width=8cm,keepaspectratio]{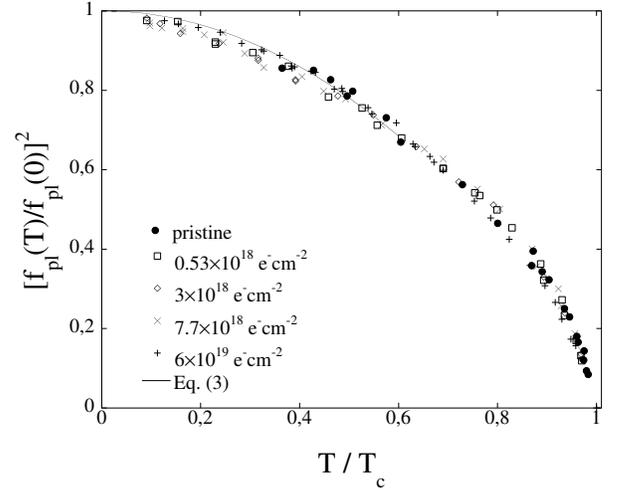}
\caption{ Square of the JPR frequency, 
normalized to its low temperature extrapolation $f_{pl}(0)$, versus reduced
temperature $T / T_{c}$. This plot is representative of the 
temperature dependence of the $c$-axis superfluid stiffness, as well 
as of the maximum $c$-axis Josephson current: $f_{pl}^{2}(T)/ f_{pl}^{2} (0)
\sim \lambda_{c}^{2}(0)/\lambda_{c}^{2}(T) \sim
\rho_{s}^{c}(T)/\rho_{s}^{c}(0) \sim j_{c}^{c}(T)/j_{c}^{c}(0)$. The 
drawn line is a fit to Eq.~(\ref{eq:IAH}) with $\Delta_{0} = 2.5 k_{B}T_{c}$.}
\label{fig:fp-scaled}
\end{center}
\end{figure}

Given the success of the IAH--model in qualitatively explaining the 
temperature- as well as the disorder dependence of the JPR data, we 
perform a direct fit of $f_{pl}^{2}(T)/f_{pl}^{2}(0)$ to 
Eq.~(\ref{eq:IAH}), shown in Fig.~\ref{fig:fp-scaled}. The only 
parameter is the value $\Delta_{0} = 2.5 k_{B}T_{c}$. This value not 
only means that $\Gamma_s>30$ K, comforting our interpretation of the
decrease of $j_{c}^c \propto T_{c}$ in terms of unitary scatterers 
induced in the CuO$_{2}$ planes, it is also remarkably close to 
characteristic energy scales found in recent
Raman scattering\cite{LeTacon2007} and STM experiments.\cite{Sacks2006} 
The first study finds that in underdoped HgBa$_{2}$CuO$_{4}$, the $B_{2g}$ Raman mode, which directly couples to 
the same low-energy nodal quasi-particle excitations that are responsible for 
the reduction of the $c$-axis superfluid density in the IAH model, is characterized 
by an energy scale $\sim 2 k_{B}T_{c}$ (related to the nodal slope 
of the gap function).\cite{LeTacon2007} We surmise that  
$\Delta_{0}$ is a closely related parameter. The second study finds that the total tunneling gap amplitude 
is determined by two energy scales, the smaller  of which, $\Delta_{\phi} 
\approx 2.8 k_{B}T_{c}$, is related to 
superconductivity.\cite{Sacks2006} It is interesting that the 
existence of a second small energy scale describing nodal quasiparticle 
exitations may well be responsible for the observed suppression of 
$j_{c}^{c}$ as function of doping (decreasing $p$ or 
$\delta$).\cite{Shibauchi2006} The proportionality of this decrease 
to the ratio of the gap maximum (at the antinode) and $T_{c}$ finds a 
logical explanation if the smaller energy scale ( $\sim 2 - 2.5 
k_{B}T_{c}$ ) is what determines the magnitude of the $c$-axis 
critical current density.

Thus, the (incoherent) interlayer assisted hopping is a feasible candidate for the 
reduction of the $c$-axis superfluid density in underdoped 
Bi$_{2}$Sr$_{2}$CaCu$_{2}$O$_{8+\delta}$: it parametrically describes 
the data, and numerical values extracted for the relevant energy scale 
determining the quasiparticle excitations is the same as found in other 
experiments. The model does have several caveats: first, there is the above--mentioned 
puzzle that it requires the temperature dependence of the $ab$--plane 
superfluid density to be explained by the lifetime effect, whereas 
the same effect does not seem to play a role in the $c$-axis 
superfluid density, other than providing the quasi-particles. The second is the identification of the defects 
in the rocksalt--like layers that are responsible for interlayer 
scattering. The model by Xiang \em et al.\rm \cite{Xiang96,Xiang98} 
requires these out-of-plane defects to be weakly scattering, with a strongly 
anisotropic potential that leaves the reflected wavevector close to 
the incident (``strong forward scattering''). Although candidates may 
be out-of-plane oxygen defects\cite{Nunner2005} or cation disorder,
the constraints imposed on the scattering potential seem very strict. 
In the end, the agreement of defect-density independence of the 
experimental $f_{pl}(T)/f_{pl}(0)$  with the IAH 
prediction\cite{Xiang96} may be completely fortuitous. The 
coincidence of the temperature dependence of $f_{pl}$ of the
pristine crystals with even the most heavily irradiated ones indicates 
that disorder plays an important role in \em all \rm  samples. Notably, we 
expect $T_{c}$ of pristine crystals to be substantially suppressed 
with respect to that of hypothetically ``clean'' underdoped 
Bi$_{2}$Sr$_{2}$CaCu$_{2}$O$_{8+\delta}$. Furthermore, impurity 
scattering is likely to completely suppress \em any \rm role of quasi-particles in  
the $c$-axis electromagnetic response of this 
compound, leaving only pair tunneling.\cite{Paramekanti2000,Ohashi2000} %

The remaining mechanism for the reduction of the superfluid density 
in the presence of pair tunneling only is that of order parameter 
phase fluctuations in the CuO$_{2}$--layers.  The effect of quantum 
phase fluctuations on the $ab$--plane and $c$-axis superfluid densities 
was examined by Paramekanti {\em et al.},\cite{Paramekanti2000,Paramekanti2002} 
who performed an analytical study of an XY model for the superconducting 
order phase $\phi$ on a two-dimensional (2D) lattice of spacing $\xi_{0}$ 
(representing the coherence length), within the self-consistent harmonic 
approximation. The authors conclude that, in contrast to thermal 
phase fluctuations, quantum phase fluctuations in quasi 2D high temperature superconductors are 
important at all temperatures. The low carrier density in these 
materials leads to inefficient screening of the Coulomb interaction 
between charge carriers, and a sizeable reduction of the magnitude of 
the $ab$--plane superfluid density (without change of its temperature 
dependence). Within this model, the JPR frequency is 
also renormalized downwards because of fluctuations of the phase in 
the layers:
\begin{equation}
    f_{pl}^{2}(T)  =  f_{pl}^{2}(0) {\mathrm e}^{  - \langle \frac{1}{2} \phi_{\perp}^{2} \rangle }  
    \approx  f_{pl}^{2}(0) \left( 1 - \frac{1}{2} 
    \langle \phi_{\perp}^{2} \rangle - \ldots \right).
\label{eq:phase-fluctuations}
\end{equation}
\noindent Paramekanti \em et al. \rm estimate the phase difference between two points separated by a 
vector perpendicular to the superconducting layers as $\langle \phi_{\perp}^{2} \rangle \approx \sqrt{
e^{2}/ 4 \pi\epsilon\epsilon_{0} \xi_{0}\varepsilon_{0}(0)s}$,\cite{Paramekanti2002}  
with $\varepsilon_{0}s = h^{2}s/ 16 e^{2} \pi \mu_{0} \lambda_{ab}^{2}(0)$ 
the in--plane phase stiffness. We observe that, given the Uemura 
relation $\varepsilon_{0}s \propto k_{B}T_{c}$,\cite{Uemura89} the resulting 
expression naturally describes the experimentally observed 
exponential depression of $j_{c}^{c}$ as function of 
doping.\cite{Shibauchi2006} However, even if one explicitly develops the 
dependence of $\varepsilon_{0}s$ in terms of the variance of the \em 
in-plane \rm phase $\langle \phi_{\parallel}^{2} \rangle \sim \langle 
\phi_{\perp}^{2}\rangle $, Eq.~(\ref{eq:phase-fluctuations}) fails to describe 
the linear $f_{pl}^{2}(T_{c})$--dependence (Fig.~\ref{fig:jc-vs-Tc}). Therefore, 
the reduction of $\rho_{s}^{c}$ with increasing disorder cannot be 
ascribed to quantum phase fluctuations only -- pair-breaking in the 
CuO$_{2}$ layers must play a significant role.

A noteworthy prediction of the quantum fluctuation scenario is that 
the temperature evolution of the $c$-axis superfluid density is 
entirely determined by that of the in-plane phase stiffness,  
\em i.e. \rm the $c$-axis and $ab$--plane superfluid densities 
follow the same dependence at low temperature. Inserting the experimental 
result $\lambda_{ab} = \lambda_{ab}(0)(1 + \beta T^{2})$ into the 
prediction of Ref.~[\onlinecite{Paramekanti2002}], one would expect 
\begin{equation}
    \frac{\partial \left[ f_{pl}(T)/f_{pl}(0) \right]^{2}}{\partial 
    (T/T_{c})^{2} } = - \frac{C_{1}}{4} \beta T_{c}^{2} \sqrt{\frac{2\pi 
    e^{2}}{4 \pi \epsilon\epsilon_{0} \xi_{0}\varepsilon_{0}(0)s}},
    \label{eq:lambda-c-vs-lambda-ab}
\end{equation} 
This is experimentally verified; taking the data of Fig.~\ref{fig:delta-lambda-plot} and  
$\lambda_{ab}(0) \approx 300$ nm,\cite{Colson2003} we find that 
Eq.~(\ref{eq:lambda-c-vs-lambda-ab}) is obeyed with $C_{1} \approx 
0.3$ ( $C_{1}$ should be order unity\cite{Paramekanti2002} ). The 
experimental independence of $f_{pl}(T)/f_{pl}(0)$ on defect density
demands that $\beta T_{c}^{2} \lambda_{ab}(0)$ is  
disorder independent. Adopting the generally accepted view that 
$\lambda_{ab}^{-2}(T)$ is described by Eq.~(\ref{eq:lifetime-ab}) 
with $\Gamma_{s}$ in the unitary limit, this would imply that $\beta T_{c}^{2} \lambda_{ab}(0) \approx 
\alpha_{ab} \lambda_{ab}(0) ( T_{c} / \Gamma )^{1/2} 
(T_{c}/\Delta_{0})^{3/2}$, and therefore that $\lambda_{ab}(0)^{-2} 
\propto T_{c} / \Gamma$. This is as yet unverified, as the 
different sizes and aspect ratios of our crystals prohibit a 
direct comparison of the absolute values of $\lambda_{ab}$.

Note that the case of screening by nodal quasiparticles was also studied 
in Ref.~[\onlinecite{Paramekanti2002}]. Then, 
\begin{equation}
    \frac{\partial \left[ f_{pl}(T)/f_{pl}(0) \right]^{2}}{\partial 
    (T/T_{c})^{2} } \approx - \frac{2 \beta 
    T_{c}^{2}}{\overline{\sigma}_{q} },
\end{equation}
where $\overline{\sigma}_{q}$ is the $ab$--plane quasiparticle sheet conductivity, 
normalized to the quantum conductivity $e^{2}/h$. This formula also describes 
the temperature dependence of the data, provided that $\overline{\sigma}_{q} \approx 3$; moreover,
the ratio $\beta / \overline{\sigma}_{q} \sim \alpha_{ab} 
\Delta_{0} (T_{c}/\Delta_{0}) m / N_{s}(0) e^{2}$ should be
disorder--independent ($N_{s}(0)$ is the quasiparticle density of states 
and $m$ the effective mass). Given the theoretical expectation 
$N_{s}(0) \sim \Gamma^{1/2}$ [\onlinecite{Preosti94}] and $\Delta_{0} \propto  
1 - \Gamma$ [\onlinecite{Sun95}], this model again fails to describe the 
reduction of the zero temperature $c$-axis superfluid density in terms of 
quantum fluctuations only.

\section{\label{sec:level5} Summary and Conclusions}

We have cross-correlated the dependence of the $c$-axis Josephson 
Plasma Resonance frequency in heavily underdoped 
Bi$_{2}$Sr$_{2}$CaCu$_{2}$O$_{8+\delta}$ on temperature and 
controlled disorder (introduced by high energy electron irradiation), and 
compared both with the behavior of the in-plane penetration 
depth. It is found that the $c$-axis critical current is depressed with 
increasing disorder strength, proportionally to the critical 
temperature $T_{c}$. Both the in--plane-- and out--of--plane 
superfluid densities follow a $T^{2}$ temperature dependence at low 
$T$. The temperature dependence of the $c$-axis response is 
independent of disorder, indicating that we are probing the superfluid 
density. The superfluid response of the pristine underdoped crystals 
is indistinguishable from that of heavily irradiated ones, suggesting 
that pristine underdoped Bi$_{2}$Sr$_{2}$CaCu$_{2}$O$_{8+\delta}$ 
commonly contains sufficient disorder in the CuO$_{2}$ 
planes for the critical temperature to be significantly suppressed with respect to 
what the $T_{c}$ of the hypothetically ``clean'' material would be. 
The dominating in-plane disorder in as-grown crystals is likely to be 
of the same kind as introduced by the electron irradiation. Apart 
from  unitary scatterers in the CuO$_{2}$ planes, this also 
encompasses the ``order parameter holes'' induced by dopant oxygen and cation 
disorder in the rocksalt-like layers.\cite{Hettler2000,Nunner2005}

The experimental data were confronted with a variety of theoretical models 
describing the reduction of $c$-axis superfluid density in terms of 
either quasiparticle dynamics or quantum fluctuations of the 
superconducting order parameter phase in the 
CuO$_{2}$ layers. We find that the quantum phase fluctuation 
description\cite{Paramekanti2000,Paramekanti2002} yields excellent 
agreement as to the experimentally observed similar temperature dependences 
of $\lambda_{ab}$ and $\lambda_{c}$, and quantitatively describes the 
temperature derivative $\partial [f_{pl}(T)/f_{pl}(0)]^{2}/\partial 
(T/T_{c})^{2}$. However, it fails to describe the dependence of the 
zero-temperature Josephson Plasma frequency on disorder strength. 

We therefore surmise that the reduction of $f_{pl}(0)$ with increasing 
disorder must be due to pair-breaking within the CuO$_{2}$ layers. 
Data for $\lambda_{ab}(T)$ and $\lambda_{c}(T,\Gamma)$ are in 
agreement with scattering in the unitary limit by the 
irradiation--induced point defects in the CuO$_{2}$--planes. Only one 
model consistently describes all aspects of the reduction of the 
$c$-axis superfluid density with temperature and disorder strength. 
This is the Impurity Assisted Hopping model of Radtke {\em et 
al.},\cite{Radtke96}  
elaborated upon by Xiang and Wheatley,\cite{Xiang96,Xiang98} and by 
Kim and Carbotte.\cite{Kim2001} The model supposes a reduction of 
$\rho_{s}^{c}(T)$ through hopping of nodal quasi-particles assisted 
by weak, highly anisotropic scattering by defects in the insulating  
SrO and BiO layers. Candidates for such impurities are out-of-plane 
oxygen defects \cite{Nunner2005} or cation disorder. From a fit of 
$f_{pl}(T)$ to the IAH model, we extract the energy scale 
$\Delta_{0} \sim 2.5 k_{B}T_{c}$ characterizing nodal quasiparticle excitations. 
This is remarkably close to the number obtained by Le Tacon {\em et 
al.} from anisotropic Raman scattering,\cite{LeTacon2007} giving 
further confidence in the IAH interpretation.

 \section*{Acknowledgements}
 This work was supported in part by the French-Japanese bilateral 
 program SAKURA, Grant No. 122313UL. 
 C.J. van der Beek wishes to thank the Department of Physics of Kyoto 
 University, where the surface resistance measurements were performed, 
 for its hospitality. We thank A.E. Koshelev for useful discussions 
 and a thorough reading of the manuscript. One of us (P.G.) was 
 partially supported by MNiSW Grant No. N202 058 32/1202.

\end{document}